\documentstyle[12pt]{article}
\begin{document}
\baselineskip 3.9ex
\def\be{\begin{equation}}
\def\ee{\end{equation}}
\def\ba{\begin{array}{l}}
\def\ea{\end{array}}
\def\bea{\begin{eqnarray}}
\def\eea{\end{eqnarray}}
\def\ub{\underbar}
\def\no#1{{\tt  hep-th#1}}
\def\nn{\nonumber}
\def\nl{\hfill\break}
\def\ni{\noindent}
\def\om{\omega}
\def\bibi{\bibitem}
\def\c#1{{\hat{#1}}}
\def\eq#1{(\ref{#1})}
\def\pgap{\vspace{1.5ex}}
\def\ggap{\vspace{10ex}}
\def\gap{\vspace{3ex}}
\def\del{\partial}
\def\o{{\cal O}}
\def\z{{\vec z}}
\def\re#1{{\bf #1}}
\def\av#1{{\langle  #1 \rangle}}
\def\S{{\cal S}}
\def\sbh{S_{\rm BH}}
\renewcommand\arraystretch{1.5}

\begin{flushright}
TIFR-TH-00/55\\
\end{flushright}
\begin{center}
\vspace{2 ex}
{\large{\bf Status of Microscopic Modeling 
of Black Holes by 
$D1-D5$ System }}$^*$\\
\vspace{3 ex}
Spenta R. Wadia\footnote{ email: wadia@tifr.res.in}\\
{\sl Department of Theoretical Physics,  
Tata Institute of Fundamental Research,}\\
{\sl Homi Bhabha Road, Mumbai 400 005, INDIA. }\\
{and\\}
{\sl Abdus Salam International Centre for Theoretical
Physics,}\\
{\sl Trieste, P.O. Box  586, I-34100, ITALY.}\\
\vspace{3ex}
$^*$ Talk given at 9th Marcel Grossmann Meeting, Rome, 
July 2000

\vspace{10 ex}
\pretolerance=1000000
\bf ABSTRACT\\
\end{center}
\vspace{1 ex}
We briefly review the microscopic modeling of black holes as bound
states of branes in the context of the soluble $D1-D5$ system.  We
present a discussion of the low energy brane dynamics and account for
black hole thermodynamics and Hawking radiation rates. These
considerations are valid in the regime of supergravity due to the
non-renormalization of the low energy dynamics in this model. Using
Maldacena duality and standard statistical mechanics methods one can
account for black hole thermodynamics and calculate the absorption
cross section and the Hawking radiation rates. Hence, at least in the
case of this model black hole, since we can account for black hole
properties within a unitary theory, there is no information paradox.
\vfill

\clearpage
\newpage

\section{\bf Black Holes, QFT and Information Puzzle}

One of the most important aspects of string theory is that gravity is
a prediction of string theory \cite{Yoneya:1974jg},
\cite{Scherk:1974mc} \cite{Callan:1985ia}.  Since string theory is
consistent with quantum mechanics (in particular it is unitary and
finite) it is widely believed that it is also a consistent theory of
quantum gravity and hence in principle should be able to resolve the
conundrums of general relativity.

One of these conundrums goes by the name of the ``Information
Puzzle'', which is intimately tied
to the fact that black holes have an event
horizon.  In the classical theory the horizon is a one way gate, in
the sense that once a particle is inside it cannot get out because of
the causal structure of the black hole space time.  In the quantum
theory  \cite{Hawking:1975sw}
black holes radiate and the radiation (at least in the
semi-classical calculation valid for large mass black holes) is
supposed to be exactly thermal%
\footnote{The causal structure
of the horizon plays an important role in this
derivation}.  The radiation is characterized 
by the Hawking temperature
\be
T_H = \frac{\hbar \kappa}{2 \pi}
\ee
$\kappa$ is surface gravity (acceleration due to gravity
felt by a static observer) at the horizon of the black hole.
For a Schwarzschild black hole: 
\be 
\kappa = \frac{1}{4 G_N M} 
\ee
Black holes are also characterized by an entropy
proportional to the area of the even horizon
\be
\S_{\it bh} = a A_h, \qquad a= \frac{c^3}{4 G_N \hbar}
\label{entropy}
\ee
The constant of proportionality $a$ is determined using 
the result from thermodynamics $TdS=dM$.
Eqn. \eq{entropy}  is the celebrated formula of Bekenstein and 
Hawking \cite{Bekenstein:1972tm}.
Hawking also gave a formula for the decay rate of a black hole in
terms of the absorption cross section $\sigma_{abs}(\omega)$: 
\be
\Gamma_H = \sigma_{abs}(\omega) (e^{\omega/T_H}-1)^{-1} \frac{d^3k}{(2
\pi)^4} 
\ee

Hawking radiation as calculated in semi-classical general relativity
is described by a mixed state. If this were exactly true it would be
in conflict with the known principles of quantum mechanics. This
conundrum is called the information puzzle.

The information puzzle would cease to exist if one could show that
Hawking radiation is similar to radiation from a standard black body
which is describable by unitary quantum mechanical evolution and the
actual wave function of the black body, although hugely complicated,
could be discerned from subtle correlations existing among the
radiated particles that come out.  It turns out to be difficult to
calculate such correlations in the case of Hawking radiation in the
standard framework of general relativity.
\cite{'tHooft:1996tq}
Such a calculation would require a good quantum theory of gravity
where controlled approximations are possible.

\section{\bf String Theory Framework for Black Holes}

String theory provides a calculable quantum theory of gravity. The
theory, as we know it, is unitary and hence it can attempt to explain
black hole thermodynamics in terms of the known principles of
statistical mechanics.
This means that in string theory a black
hole should be described by a density matrix:
\bea
\label{density}
\rho & = & {1\over \Omega }\sum_{i}|i\rangle \langle i|  \nn \\
S & = & \ln{ \Omega}
\eea
where $|i\rangle $ is a micro-state.

In such a framework Hawking radiation is no different from the
radiation emitted by burning a piece of wood. The thermal description
is a useful gross description in which one averages over a large
ensemble of states. In such a framework the Bekenstein-Hawking formula
is the same as Boltzmann's formula and the Bose factor in the formula
for the decay rate corresponds to the statistical average of the
occupation number at the frequency $\omega$.

\gap

\noindent{\bf Ingredients of String Calculation}

\gap

We enumerate the basic ingredients we need to do the string theory
calculation:

\begin{enumerate}
\item We need a microscopic model of the black hole 
and the effective Lagrangian of the low energy excitations
of this model at strong coupling.
\item In order to calculate the Hawking process we need
the interaction of the effective degrees of freedom with the
supergravity modes which are radiated by the black hole.
\item Once we can calculate transition amplitudes in a unitary theory
between black hole states we can derive black hole thermodynamics 
using the micro-canonical or canonical ensemble.
\end{enumerate}

The model which allows string theory calculation under controlled
approximations, is the near extremal 5-dim. black hole of type II-B
string theory compactified on a 4-torus (or $K_3$) \cite{Str-Vaf96}.
This black hole has a very small temperature, so that the thermal
wavelength is much larger than the typical gravitational radius of the
black hole.

\section{\bf The Near Extremal Black Hole}

We now elaborate a bit more about the black hole we are modeling.  The
near extremal black hole solution of II-B string theory compactified
on $T^4\times S^1$ preserves none of the original 32 supersymmetries
of the type IIB theory. It has a non-zero Hawking temperature and a
positive specific heat (unlike the Schwarzschild black hole).  We want
to model the thermodynamics of this solution in string theory.

The solution involves the metric, the dilaton and the Ramond-Ramond
2-form $C^{(2)}$. The coordinates $x^{i}, i=1,2,3,4$ are non-compact.
$x^5$ is periodically identified with period $2\pi R_5$ and directions
$x^6, \ldots,x^9$ are compactified on a torus $T^4$ of volume $V_4$,
$R_5 \gg (V_4)^{1/4}$.

This solution is parameterized by six independent quantities: $r_1,
r_5, r_0, \sigma, R_5$ and $V_4$ (notations
defined in \cite{Mal-Str96}). These are related to the number
$Q_1$ of D1-branes, $Q_5$ of D5-branes and Kaluza-Klein momentum N to
be distributed in both directions around the $x_{5}$ direction. $r_0$
is the non-extremality parameter.
 
The entropy and mass of the black hole are well known. The near
extremal black hole has a small Hawking temperature and unlike the
Schwarzschild black hole a positive specific heat.

The classical solution is relevant in the quantum theory only if
quantum loops are suppressed: $g_{s}\rightarrow 0$. Hence we require
$g_sQ_1$, $ g_sQ_5$ to be held fixed and we are dealing with the large
$Q_1, Q_5$ limit. For a macroscopic black hole the horizon area is
much larger than the string length $l_{s}=\sqrt{\alpha ^{'}}$.  This
implies $g_sQ_1>>1, \; g_sQ_5>>1, \; g_s^2N >>1$.  Since, $g_sQ_1$, $
g_sQ_5$ correspond to the effective open string coupling constants, a
macroscopic black hole exists at strong coupling!

\section{\bf Absorption Cross Section, Decay Rate}

Using the black hole solution one can calculate by solving the wave
equation the absorption cross section for various particles. The
simplest one to do are the so called minimal scalars because they only
couple to the background Einstein metric and their wave equation is
\be D_\mu \del^\mu \varphi =0 \ee 
The s-wave absorption cross section
is given by \be \sigma_{abs} = 2 \pi^2 r_1^2 r_5^2 \frac{\pi
\omega}{2} \frac{\exp(\omega/T_H)-1}{(\exp(\omega/2T_R)-1)
(\exp(\omega/2T_L)-1)} \ee In the $\omega\to 0$ limit, one gets \be
\label{abs}
\sigma_{abs} = A_h
\ee
where $A_h$ denotes the area of the event horizon.
The decay rate is 
\be
\Gamma_H = \sigma_{abs} (e^{\omega/T_H}-1)^{-1} 
\frac{d^4k}{(2 \pi)^4}
\ee
The absorption cross section of higher partial waves vanishes in 
the $\om\to 0$ limit.

\section{\bf Microscopic Model: D1-D5 System}

We begin with the theory of the $Q_5$ $D_5$ branes along the compact
coordinates $x_i, i=5,6,7,8,9$.  The low energy degrees of freedom of
this system are described by an $N=2$ $U(Q_5)$ gauge theory in 6
dimensions.

In this gauge theory the configurations which break the 16
supersymmetries to 8 are instantons.  Along the $x_5$ direction this
is a string like configuration and in fact it is to be identified with
$Q_1$ $D_1$ branes, if the instanton charge is $Q_1$.  The instantons
are characterized by moduli whose variation does not change the action
of SYM$_6$. Promoting these moduli to slowly varying functions of
$x_5,t$ we obtain the motions of the $D_1$ branes inside the
5-branes. These represents the low lying collective modes of the
$D_1,D_5$ system.
It turns out that the moduli
space ${\cal M}$, of instantons on $T^4$, is the Hilbert Scheme of the
symmetric product $(\tilde{T}^4)^{Q_1 Q_5}/S(Q_1Q_5)$. ($\tilde{T}^4$
can be different from the compactification torus $T^4$.)

Our attitude will be to consider the sigma model on ${\cal M}$, as a
resolution of the sigma model on the orbifold $(\tilde{T}^4)^{Q_1
Q_5}/S(Q_1Q_5)$. ${\cal M}$ is a hyper-kahler manifold and hence one
can define a $N=(4,4)$ SCFT. We can explicitly construct the $N=(4,4)$
orbifold SCFT. The 4 marginal operators of the SCFT are the blowing up
modes of the orbifold.
Before we summarize the SCFT we list a few points regarding the
validity of our considerations in the strong coupling region where
$g_sQ_1Q_5\gg 1$.
\begin{itemize}
\item The instanton equation is derived as a condition from
supersymmtry and it is independent of the coupling constant and 
$\alpha '$.
\item The moduli space and the corresponding sigma
model does not receive any corrections in the string coupling.
because the hypermultiplet moduli space does
not get renormalized by the interactions.
This fact is crucial because it says that the SCFT that we found
at weak coupling is valid at strong coupling. Hence we can use it
to make comparisons with supergravity calculations of the entropy and 
temperature of the black
hole and the Hawking rates corresponding to particles which are in
the short multiplets of the $N=4$ superconformal algebra.
\end{itemize}

\section{\bf N=4 SCFT on ${\rm Sym}(\tilde{T}^4)$}

The $N = (4,4)$ SCFT on ${\rm Sym}(\tilde{T}^4)$
is described by the free Lagrangian
\be
\label{free}
S = \frac{1}{2} \int d^2 z\; \left[\del
x^i_A \bar\del x_{i,A} + 
\psi_A^i(z) \bar\del \psi^i_A(z) + 
\widetilde\psi^i_A(\bar z) \del \widetilde \psi^i_A(\bar z) 
 \right]
\ee
Here $i$ runs over the $\widetilde{ T^4}$ coordinates
1,2,3,4 and $A=1,2,\ldots,Q_1Q_5$ labels various copies
of the four-torus. The symmetric group $S(Q_1Q_5)$
acts by permuting the copy indices.

The central charge of the SCFT is $c=6Q_1Q_5$.
The N=4 algebra contains the global supergroup 
$SU(1,1|2)\times SU(1,1|2)$ which contains the bosonic subgroup 
$SL(2,R)\times SU(2)_R$. The subscript R indicates the R-symmetry.
The theory also has an additional global symmetry group $SO(4)_I$.

$SU(1,1|2)$ has 8 real supercharges and hence the total number of
supercharges is 16. 
This is in contrast to the fact that the blackhole
solution even with KK charge $N=0$ had only 8 SUSYS! 
This puzzle was
resolved by Maldacena 
\cite{malda-dual} who showed that the relevant supergravity
solution is the so called near horizon geometry viz. 
$AdS_3\times S^3\times \tilde{T}^4$.
The conformal and R-symmetries of the SCFT become isometries of the
near horizon geometry.

\gap

\section{\bf Matching SCFT Operators and SUGRA Moduli using
Maldacena duality}

The duality conjecture of Maldacena \cite{malda-dual} as far as the
symmetries are concerned states that the $SU(1,1|2)\times SU(1,1|2)$
isometries of the near horizon geometry are matched with the global
symmetries of the ${\cal N}=(4,4)$ SCFT on ${\cal M}$. Further the
$SO(4)_I$ algebra of $T^4$ is identified with $SO(4)_I$ algebra of
$\tilde{T}^4$. It is important to note that even though $SO(4)_I$ is
not a symmetry of $T^4$ it is useful to classify the low energy
states.

The representations of ${\cal N}=(4,4)$ can be classified in terms of the
chiral primary operators which are specified by 
$h=j$ ({\it i.e}, conformal dimension = spin). 
Similarly $\bar h = \bar j$.
The chiral primary is denoted by $(\bf{2h +1}, \bf{2h'+1})_S$
For each chiral primary there corresponds a short multiplet obtained 
by the action of the global supercharges 
$G^{1\dagger }_{-1/2}, G^2_{-1/2}$.

All the chiral primaries for the orbifold conformal field theory
(OCFT) on ${\cal M}$ can be explicitly constructed by the product of
the chiral primaries corresponding to the cohomology of the diagonal
$\tilde{T}^4$ (the sum of all copies of $\tilde{T}^4$) and the various
k-cycle chiral primaries. The latter are in one-to-one correspondence
with the cyclic subgroups of $S(Q_1Q_5)$ which are characterized by
the length $k$ of the cycle. $k=1,2,...Q_1Q_5$. 
The conformal dimension and spin of the 
chiral primary is $((k-1)/2, (k-1)/2)$. Note that $k_{max}=Q_1Q_5$.
This upper bound is called the stringy exclusion principle.

The SCFT under consideration has 5 (2,2) short multiplets. These
chiral primaries constitute 20 relevant operators of the SCFT.  
In the OCFT these operators belong to the the cyclic subgroup of
length $2$. $4 (2,2)$ short multiplets come from the untwisted sector
and one from the $Z_2$ twisted sector.
The corresponding top components of the $(2,2)$ short multiplets are
the 20 marginal operators. The 4 marginal operators that come from the
$Z_2$ twisted sector are the 4 blowing up modes of the OCFT. One can
show that the 20 marginal operators of the SCFT define the
Zamolodchikov metric of the coset $\frac{SO(4,5)}{SO(4)\times SO(5)}$.
The number of marginal operators is a topological invariant. This is
because the number of chiral primaries with $(j_R, \tilde
j_R)=(m,n)$ is the Hodge number $h_{2m,2n}$ of the target space ${\cal
M}$ of the SCFT. 
As far as the quantum numbers of the 20 marginal operators are
concerned a distinction can be made only on the basis of their
$SO(4)_I$ quantum numbers (See table below). 

The SUGRA moduli in the near horizon geometry 
are classified using the formula:
\be
h+\bar h = 1+ \sqrt {1+m^2}
\ee
and the global symmetry $SO(4)= SU(2)_I\times \widetilde{SU(2)_I}$
We see that marginal operators 
correspond to massless excitations in SUGRA.
The 20 massless scalars (see table below) form the coset 
$\frac{SO(4,5)}{SO(4)\times SO(5)}$.
Once again the distinct quantum numbers come from the global 
$SO(4)$ symmetry.

The matching of the marginal operators of the SCFT and the SUGRA moduli 
is summarized in the table below:
\bea
\begin{array}{llc}
\mbox{Operator} & \mbox{Field}  &  
SU(2)_I\times \widetilde{SU(2)}_I \\
\del x^{ \{ i }_A(z) \bar{\del}x^{ j\} }_A (\bar z) -1/4\delta^{ij}
\del x^{k}_A \bar{\del}x^k_A
& h_{ij} -1/4\delta_{ij} h_{kk} & (\bf 3, \bf 3 ) \\
\del x^{[i}_A(z) \bar{\del}x^{j]}_A (\bar z) 
& b'_{ij}
& (\bf 3, \bf 1 ) +
(\bf 1, \bf 3)  \\ 
\del x^i_A(z) \bar{\del}x^i_A (\bar z) 
& \phi
&( \bf 1, \bf 1 ) \\
{\cal T}^1 & b^+_{ij} & (\bf 1, \bf 3 ) \\
{\cal T}^0 & a_1C_0 + a_2C_{6789} & (\bf 1, \bf 1 )
\end{array}
\eea
${\cal T}^1$ and ${\cal T}^0$ are twisted sector operators.  $b^+_{ij}$ is
the self dual part of $B_{NS}$. It is the modulus that leads to stable
(non-marginal) bound states. $C_0$ and $C_{6789}$ are Ramond fields.  In
order to make a precise matching of the above we have related the
blowing up modes of the SCFT comming from the twisted sector with the
stabilizing moduli in supergarvity.

\section{\bf Maximally Twisted Sector and Black Hole Hilbert Space}

We now investigate the Hilbert space of the black hole states.
The longest cyclic subgroup of $S(Q_1Q_5)$ has length $Q_1Q_5$ and
leads to the the maximally twisted sector of the orbifold SCFT, which
is characterized by a chiral primary $((Q_1Q_5-1)/2, (Q_1Q_5-1)/2)$.
Since this belongs to the longest cyclic subgroup there are no chiral
primaries of higher spin. 
The presence of this twist field leads to twisted boundary conditions
for the basic coordinates of the orbifold. 
\be
\label{chp4:bc}
X_A (e^{2\pi i}z,e^{-2 \pi i} \bar{z})=X_{A+1} (z,\bar{z})
\ee
This implies that the momentum $n_L, n_R$ in the twisted sector is quantized in units of $1/(Q_1Q_5)$, and hence the momentum quantum number can go upto an integer multiple of $(Q_1Q_5)$.

The bh micro-states are defined by the level conditions
\be
L_0=N_L \;\;\;\; \bar{L}_0=N_R
\ee
It reflects the fact that the general non-extremal black hole
will have Kaluza-Klein excitations along both the directions on the
$S^1$.

The leading order (in large $Q_1Q_5$) entropy formula corresponding 
to these level conditions is
\be
S = 2 \pi \sqrt{n_L} + 2 \pi \sqrt{n_R}
\ee
$n_L=Q_1Q_5N_L$ and $n_R=Q_1Q_5N_R$.
This also turns out to be the contribution to the entropy from the
maximally twisted sector. Hence the leading contribution to the black
hole entropy comes from the twisted sector. Hence we can assert that
to leading order in $Q_1Q_5$, the black hole micro-states reside in
the maximally twisted sector.
The entropy formula readily enables a calculation of the Hawking
temperature which agrees with SUGRA. The Hawking temperature is
independent of the string coupling!

\section{\bf Hawking Radiation}

Absorption cross-section of a supergravity fluctuation $\delta \phi$
is related to the thermal Green's function of the corresponding
operator ${\cal O}$ of the ${\cal N} =(4,4)$ SCFT on the orbifold
${\cal M}$ .

The absorption of a quantum $\delta\bar\phi = \kappa_5 e^{-ipx}$
corresponding to the operator ${\cal O}$ is calculated using the
Fermi's Golden Rule and is related to the discontinuity of the thermal
Green's function ${\cal G} (-i\tau, x)$ of the operator ${\cal O}$.
The Green's function ${\cal G}$ is determined by the two-point
function of the operator ${\cal O}$. This is in turn determined by
conformal dimension $(h, \bar{h})$ of the operator ${\cal O}$ and the
normalization ${\cal C}_{\cal O}$ of the two-point function.  
\def\CO{{\cal C}_{\cal O}}
\be
\sigma_{abs} =\frac{\mu^2\kappa_5^2 L}{{\cal F}} \int dt\;dx ({\cal G}
(t-i\epsilon , x) - {\cal G} (t+i\epsilon, x) ) 
\ee 
Hence 
\bea
\sigma_{abs}&=& \frac{\mu^2\kappa_5^2 L {\cal C_O}}{{\cal F}} \frac{
(2\pi T_L)^{2h -1} (2\pi T_R)^{2\bar{h} -1} } { \Gamma(2h)
\Gamma(2\bar{h}) } \frac{ e^{\beta\cdotp p/2} - (-1)^{2h + 2\bar{h}}
e^{-\beta\cdotp p/2} }{2} \\ \nonumber &\;&\left| \Gamma (h +
i\frac{p_{+}}{2\pi T_L} ) \Gamma (\bar{h} + i\frac{p_{-}}{2\pi T_R} )
\right|^2
\label{temp}
\eea

How does one fix the coefficient $\CO$
within the microscopic
theory?  Presently we do not know how to do this. All that one can say
is that this coefficient which is the normalization of the SCFT
operators undergoes no renormalization and it is independent of the
string coupling. We can determine this number by using the AdS/CFT
correspondence.

The formula \eq{temp} can be applied to the minimal scalars and the
coefficient $\CO$  can be fixed by matching the zero temperature
2-point function of the minimal scalar $h_{67}$ corresponding to the
fluctuation of the metric of $T^4$. $\CO$ is the same for all
the minmal scalars because the corresponding operators of the SCFT
define the Zamolodchikov metric of the coset
$\frac{SO(4,5)}{SO(4)\times SO(5)}$.  $\CO$ is then the
relative normalization between the Zamolodchikov metric and the metric
in supergravity of the same coset.

In this way we find the minimal scalar absorption cross section.
\be
\sigma_{abs}=
2 \pi^2 r_1^2 r_5^2 \frac{\pi \om}{2} 
\frac{\exp(\om/T_H)-1}{(\exp(\om/2T_R)-1) (\exp(\om/2T_L)-1)}
\ee
Thus the SCFT calculation and the supergravity calculation of the
absorption cross-section agrees exactly with the semiclassical result.

\vspace{2ex}

\noindent{\sl Moduli and Hawking Radiation}

\vspace{2ex}

In the semiclassical calculation it is clear that absorption cross
section is independent of the presence of vevs of the massless fields.
It is possible to show that
\cite{Dav-Man-Wad99} the same is true in the SCFT. 

\section{What's new}

It needs a lot of work to be able to
correctly calculate the absorption
crossection and the Hawking rates which agree with the semi-classical
supergravity calculations. The string theory calculations were
originally done in \cite {Dha-Man-Wad96,Das-Mat96} and were based on a
model that was physically motivated by string dualities.  In
particular the calculation in \cite{Das-Mat96} based on the DBI action
reproduced even the exact coefficient that matched with the
semi-classical answer for the absorption cross section of the minimal
scalars.  However this method did not work when applied to the fixed
scalars \cite {Kra-Kle97}.  This fact was very discouraging because it
meant the absence of a consistent starting point for string theory
calculations. The discovery of Maldacena \cite {malda-dual} finally
enabled the string theory calculations \cite {Dav-Man-Wad98} because
it was able to make a precise connection of the near horizon geometry
with the infra-red fixed point theory of brane dynamics.

\section{Open problems}

Let us conclude by stating some interesting problems:

\begin{itemize}
\item How does one formulate the effective long wave length theory of the 
non-supersymmetric black holes? 

\item How does one derive space-time from brane theory? In particular
is there a way of deducing $AdS_3\times S^3$ (the infinitely stretched
horizon) as a consequence of brane dynamics?  The method of co-adjoint
orbits applied to the SCFT is a promising approach. And what about the
black hole horizon itself? These questions are intimately tied to
explaining the geometric Bekenstein-Hawking formula or in other words
understand the holographic principle \cite{tHooft}.

\item The D1/D5 system has relevant perturbations. It would be
interesting to understand to study the holographic renormalization
group in this situation. What is the end point of the RG flow?

\end{itemize}

\vspace{2ex}

{\bf Acknowledgments}

\vspace{2ex}

I would like to thank Gautam Mandal for a critical reading
of the manuscript.

\end{document}